\documentclass[12pt]{article}
\begin{document}
\begin{center}
{\large{Deformed Special Relativity and Deformed Symmetries \\
in a Canonical Framework }}
   \vskip 2cm

Subir Ghosh\\
\vskip .3cm
Physics and Applied Mathematics Unit,\\
Indian Statistical Institute,\\
203 B. T. Road, Calcutta 700108, India\\
\vskip .3cm
And\\
\vskip .3cm
Probir Pal\\
Physics Department,\\
Uluberia College,\\
Uluberia, Howrah 711315,  India.\\
\end{center}
\vskip 3cm
{\bf Abstract:}\\
In this paper we have studied the nature of kinematical and
dynamical laws in $\kappa $-Minkowski spacetime from a new
perspective: the canonical phase space approach. We discuss a particular form of $\kappa$-Minkowski phase space algebra that yields  the $\kappa$-extended finite Lorentz
transformations derived in \cite{kim}. This is a particular form
of a Deformed Special Relativity  model that admits a modified
energy-momentum dispersion law as well as noncommutative
$\kappa$-Minkowski phase space. We show that this system can be
completely mapped to a set of phase space variables that obey
canonical (and {\it{not}} $\kappa$-Minkowski) phase space algebra
and Special Relativity Lorentz transformation (and {\it{not}}
$\kappa$-extended Lorentz transformation). The complete set of deformed symmetry generators are constructed that obeys an unmodified closed algebra but induce deformations in the symmetry transformations of the physical $\kappa$-Minkowski phase space variables. Furthermore, we demonstrate the usefulness and simplicity of this approach  through a number of phenomenological
applications both in classical and quantum mechanics. We also
construct a Lagrangian for the $\kappa$-particle.

\newpage

{\bf{Introduction:}}\\
\vskip .2cm Evidence (see \cite{am1} for discussion and
references) of ultra-high energy cosmic ray particles that violate
the Greisen-Zatsepin-Kuzmin bound have compelled theorists to
generalize the conventional energy-momentum dispersion law of
particles,
\begin{equation}
p^2=m^2,  \label{sr}
\end{equation}
based on principles of Special theory of Relativity (SR). The
extension requires another observer independent dimensional
parameter, apart from $c$, the velocity of light. The second
parameter $(\kappa )$ is expected to be related to Planck energy.
Based on this idea Amelino-Camelia \cite{am} has pioneered an
extended form of SR, popularly known as Deformed (or Doubly)
Special theory of Relativity (DSR). It is very important to
emphasize that the fundamental tenet of SR, {\it{i.e.}}
equivalence of inertial frames of reference is kept intact in DSR whereas the Poincare algebra  of SR is elevated to quantum (Hopf) $\kappa$-Poincare algebra in DSR. The effect
of $\kappa $ appears in the explicit structures of Lorentz
transformations (LT) in DSR, referred to here as $\kappa$-LT,
which are non-linear for momenta and momenta-dependent for coordinates. Also DSR fully reduces to SR for
energies much smaller that Planck energy, (or equivalently in the
limit $\kappa \rightarrow \infty $).

DSR is intimately connected to another area of topical interest:
Non-Commutative (NC) geometry \cite{sn,sw}, where canonical
Poisson Bracket structure of phase space variables is replaced by
a more extended algebra. This is indeed satisfying because
existence of a fundamental length scale ($\sim$ Planck length) is
needed to deal with quantum gravity ideas \cite{planck} and  Black
Hole physics \cite{fred} in a sensible way. This additional length
scale can be implemented by an NC spacetime via generalized
uncertainty relation \cite{kem}. Exploiting the notion of duality
in the context of Quantum Group ideas \cite{dsr2,mag,kow} or from
a physical point particle Lagrangian perspective \cite{sg1}, it
has been demonstrated
 that a DSR dispersion law is the Casimir
invariant of a particular $\kappa $-deformed Poincare algebra and
the latter is connected to an NC phase space in a unique way.   It is
important to note that in these examples the NC extension of the spacetime
algebra is operatorial in nature, 
$$\{x_\mu ,x_\nu \}=\theta_{\mu\nu}(x_{\alpha},p_{\alpha}),$$
($p_\alpha $ being the momentum),
in contrast to NC spaces of
recent interest in High Energy Physics, where the extension is not
operatorial \cite{sw},
$$\{x_\mu ,x_\nu \}=\theta_{\mu\nu}.$$

In the present paper we will dwell on the NC phase space aspect of
DSR. Our approach \cite{sg1} is more familiar to a physicist than
the more mathematically oriented previous studies
\cite{dsr2,mag,kow}. We will focus on a particular form of NC
spacetime, known as $\kappa$-Minkowski spacetime (and the
associated phase space). In the sense of classical Poisson Brackets, the NC $\kappa$-Minkowski spacetime is defined as, 
\begin{equation}
\{x^i,x^0\}=\frac{x^i}{\kappa} ~;~~
\{x^i,x^j\}=0.
\label{ok}
\end{equation}
It should be mentioned that, even if one imposes the restrictions that Jacobi
identities have to be maintained and that the structure should
reduce to canonical algebra for $\kappa \rightarrow \infty $, the
full $\kappa$-NC phase space algebra  is not uniquely determined. There are
distinct (and possibly inequivalent) representations that are connected by non-linear transformations \cite{kow}. The particular $\kappa$-NC phase space that we will use here was first studied in \cite{granik} and was further developed in \cite{mig} (in a restricted set up of $1+1$-dimensional toy model). In fact this phase space can be extracted from very general deformations considered by Lukierski et.al. (see the last reference in \cite{dsr2}). In \cite{granik,mig} the idea of a suitable definition for velocity of a particle, with a Magueijo-Smolin form \cite{mag} of modified dispersion relation,
\begin{equation}
p^2= m^2[1-\frac{E}{\kappa}
]^2,  \label{0m}
\end{equation}
$E$ being the particle energy, led to this NC phase space. Interestingly enough, this same phase space was embedded in a more general structure in an earlier work of one of us \cite{sg1} where it was induced in a specific gauge that was used to fix the reparameterization invariance. (Similar framework has been used in various other contexts \cite{others} to generate NC phase space.) We will follow
the method used in \cite{bru} and obtain the expressions for
{\it{finite}} $\kappa$-LTs.  We will start from the
infinitesimal $\kappa$-LTs induced by Lorentz generators in
$\kappa$-space by explicit usage of the $\kappa$-NC algebra
proposed here. The finite $\kappa$-LTs will be composed out of
them by integration \cite{bru}. These specific $\kappa$-LTs were first derived   in \cite{kim} from a different perspective. Also they were generated in \cite{mig} in the symplectic framework. We, on the other hand, will demonstrate that compatibility with the $\kappa$-NC phase space and associated $\kappa$-LTs  necessarily lead to  the generalized MS dispersion
relation  \cite{mag},
\begin{equation}
p^2=m^2[1-\frac{(\eta p)}{\kappa} ]^2 = m^2[1-\frac{E}{\kappa}
]^2,  \label{m}
\end{equation}
now written in a covariant form with  $\eta^0=1,\vec \eta =0$. We
also show that the SR LT invariant interval has to be replaced by
an expression that is DSR or $\kappa$-LT invariant \cite{mig}.

In generalizations of Relativity theory in the form of introducing more than one observer independent scales, ({\it{e.g.}} in DSR there are two scales), generally the schemes  are dominated by algebraic constructions based on the quantum (Hopf) $\kappa$-Poincare algebra \cite{dsr2,kow}. This is a deformation of the normal Poincare algebra of SR that is applicable in conventional phase space. {\footnote{The construction of quantum field theories on a canonical NC space, (where $\theta_{\mu\nu}$ is a constant in $\{x_\mu ,x_\nu \}=\theta_{\mu\nu}$), has received a great impetus from the works of  \cite{chi}, that demonstrates the invariance of the theory under twisted Poincare algebra. This means that the representation content of NC $x_\mu $ remains the same as there commutative counterpart. This assertion is very important from phenomenological model building point of view \cite{chi}.}} $*$-Product, an essential tool in the construction of quantum field theory in NC space, for the $\kappa$-NC spacetime has appeared in \cite{*-kappa}. This aspect of our model will appear elsewhere \cite{future} and in the present work we will provide a thorough analysis of the deformed symmetry generators and their effects on the degrees of freedom.

Now comes the major part of our work where we introduce a
mapping between $(x_\mu ,p_\mu )$ - the physical $\kappa$-Minkowski NC
phase space and $(X_\mu ,P_\mu )$ - a completely canonical phase
space that  obey normal Poisson Bracket algebra. We also show that
$(X_\mu ,P_\mu )$ obey SR LTs  whereas $(x_\mu ,p_\mu )$ are ruled
by  $\kappa$-LTs. Similar ideas have appeared before in $1+1$-dimensions in \cite{mig} and also in 
partially restricted ways in either coordinate space \cite{der} or in
momentum space \cite{hos}. However,  the full canonical  phase
space, as has been constructed in the present work  via the
mapping, becomes a  powerful tool. The idea of covariant phase space has also been applied \cite{arz} in some $\kappa$-extended field theory models to identify the conserved quantities.

On the one hand, existence of the associated canonical phase space
is very intriguing and requires serious thinking, whereas on the
other hand, it has a great practical utility in describing
$\kappa$-LT kinematics and dynamics. It will also play an
important role in quantization of the $\kappa$-particle. The idea
is to write down SR or quantum relations in terms of $(X_\mu
,P_\mu )$ and then use the inverse map to get corresponding
relations in terms of $(x_\mu ,p_\mu )$ variables, which (although
NC) are the relevant physical degrees of freedom. However, we also
show that one has to be careful in extending SR kinematic
relations to DSR relations since for interacting particles, the
dynamic behavior becomes entwined with kinematics and isolated
kinematic identities (as in SR case) are allowed only for free
particle. Incidentally, even for the free $\kappa$-particle, the
equation of motion (Newton's law) $\dot p_\mu =0$ is true but it is not at all
obvious and we prove this from an action principle for the
$\kappa$-particle.  In the quantum case, we show from
explicit examples that $\kappa$-effects are not observable in
non-relativistic situations.

 We provide a first
order phase space Lagrangian for the $\kappa$-point particle and
its geometric counterpart in coordinate space that has a higher
derivative structure. (Higher derivative terms have also appeared
in other DSR particle model \cite{gir1}.) Our model includes the
two stand out features of DSR systems: NC  spacetime and the
modified dispersion relation, (taking $\kappa$-Minkowski spacetime
and MS dispersion law as an example){\footnote{The present model is a minimal construction in the sense that the previous model constructed  by one of us \cite{sg1} had a more elaborate phase space algebra.}} . As we will show, here also
the natural starting point is the normal relativistic point
particle model in terms of $(X_\mu ,P_\mu )$ which is  rewritten
in terms of NC coordinates $(x_\mu ,p_\mu )$. We believe that the usage of this
equivalent canonical framework, described extensively in the present
work, to construct the deformed symmetry generators and the MS point particle Lagrangian, is new.

The paper is organized as follows: In Section II we introduce the
 $\kappa$-Minkowski phase space, the infinitesimal Lorentz
generators and construct the finite $\kappa$-LTs for the phase
space coordinates, {\it{i.e.}} $x_\mu $ and $p_\mu $. The
$\kappa$-LT invariant MS dispersion law and coordinate interval
are also identified. Section III is the main content of our work.
In this section the canonical $(X_\mu ,P_\mu )$ phase space is
constructed out of $(x_\mu ,p_\mu )$, the $\kappa$-NC phase space.
The SR LT of the set $(X_\mu ,P_\mu )$ are also shown explicitly.
In Section IV we provide a thorough analysis of the set of deformed symmetry generators, comprising of generators of  Translation, Rotation, Dilatation and Special Conformal Transformation, using the canonical approach. In Section V, we discuss kinematics in the $\kappa$ DSR context.
Section VI is devoted to the construction of an action for the free
$\kappa$-particle that is the starting point of all dynamical
studies. The quantization problem is briefly touched upon in
Section VII. The paper ends with a conclusion in Section VIII.

 \vskip .4cm
{\bf{II: Non-linear $\kappa$-Lorentz Transformations}}\\
\vskip .2cm Let us start by introducing the $\kappa$-Minkowski
phase space. We are in the classical framework and will interpret
the phase space algebra as Poisson brackets. The general consensus
is to refer a phase space as $\kappa$-Minkowski if it contains the
following sector,
\begin{equation}
\{x^i,x^0\}=\frac{x^i}{\kappa} ~;~~
\{x^i,x^j\}=0~;~~\{x^i,p^j\}=-g^{ij}~;~\{p^\mu,p^\nu\}=0.
\label{ko}
\end{equation}
Our metric is $diag~g^{00}=-g^{ii}=1$ and $\kappa $ is the  NC
parameter. Rest of the phase space algebra is given
below,
\begin{equation}
 \{x^0,p^i\}=p^i/\kappa ~;~\{x^i,p^0\}=0~;~\{x^0,p^0\}=-1+p^0/\kappa.
\label{2}
\end{equation}
The above is rewritten in a covariant form,
$$
\{x_\mu ,x_\nu \}=\frac{1}{\kappa}(x_\mu \eta_{\nu}-x_\nu
\eta_{\mu }),$$
\begin{equation}
\{x_{\mu},p_{\nu}\}=-g_{\mu\nu}+\frac{1}{\kappa}\eta_{\mu}p_{\nu},~~\{p_{\mu},p_{\nu}\}=0,
 \label{03}
\end{equation}
where $\eta _0=1,\eta _i=0$. This algebra appeared in \cite{granik} and partially in \cite{mig}. Detailed studies of similar types of algebra are provided in \cite{kow}.  This algebra has
emerged before in an earlier work of one of us \cite{sg1} where it was
embedded in a more general algebra. For $\kappa \rightarrow \infty $
one recovers the normal canonical phase space.

The angular momentum is defined in the normal way as,
\begin{equation}
J_{\mu\nu }=x_\mu p_\nu -x_\nu p_\mu . \label{j}
\end{equation}
This is motivated by the fact that spatial sector of $\kappa$-NC algebra in (\ref{03}) remains unaffected.
Furthermore,  using (\ref{03}) one can check that the Lorentz algebra is
intact,
\begin{equation}
\{J^{\mu\nu },J^{\alpha\beta }\}=g^{\mu\beta }J^{\nu\alpha
}+g^{\mu\alpha }J^{\beta \nu}+g^{\nu\beta }J^{\alpha\mu
}+g^{\nu\alpha }J^{\mu\beta }. \label{51}
\end{equation}
 However,
Lorentz transformations of $x_\mu$ and $p_\mu $ are indeed
affected,
\begin{equation}
\{J^{\mu\nu},x^\rho \}=g^{\nu\rho}x^\mu-g^{\mu\rho}x^\nu
+\frac{1}{\kappa } (p^\mu\eta^\nu -p^\nu \eta^\mu )x^\rho ~;~
\{J^{\mu\nu},p^\rho \}=g^{\nu\rho}p^\mu -g^{\mu\rho}p^\nu
-\frac{1}{\kappa } (p^\mu\eta^\nu -p^\nu \eta^\mu )p^\rho .
\label{52}
\end{equation}
Notice that the extra terms appear only for $J^{0i}$ and not for
$J^{ij}$ so that only boost transformations are changed.

From now on we will use the $(x,y,x,t)$ notation (instead of the
covariant one), which is more suitable for comparison with
existing results. We define the infinitesimal transformation of a
generic variable $O$ by,
\begin{equation}
\delta O=\{\frac{1}{2}\omega_{\mu\nu}J^{\mu\nu},O\}, \label{l}
\end{equation}
and only the parameter $\omega_{0x}=\delta u$ is non-vanishing.

Let us start with the Lorentz transformation for energy-momentum
vector $(E,p_x,p_y,p_z)$. The above considerations yield the
following differential equations \cite{bru},
\begin{equation}
\frac{dE}{du}=-p_x+\frac{Ep_{x}}{\kappa};~~\frac{dp_x}{du}=-E+\frac{p^{2}_{x}}{\kappa};~~\frac{dp_y}{du}=\frac{p_{y}p_{x}}{\kappa};~~\frac{dp_z}{du}=\frac{p_{z}p_{x}}{\kappa}.
\label{l1}
\end{equation}
We rewrite the $E$ and $p_x$ equations as,
\begin{equation}
\frac{d(E+p_x)}{du}=-(E+p_x)+\frac{p_{x}}{\kappa}(E+p_x);~\frac{d(E-p_x)}{du}=(E-p_x)+\frac{p_{x}}{\kappa}(E-p_x),
\label{l2}
\end{equation}
which reduces to,
\begin{equation}
\frac{d}{du}[ln(\frac{(E+p_x)}{(E-p_x)})]=-2. \label{l3}
\end{equation}
Introducing the initial and final respectively as  $E,p_x$ for
$u=0$ and $E',p'_x$, we obtain the following relation,
\begin{equation}
E'(1-Ae^{-2u})=-p'_x(1+Ae^{-2u}), \label{l4}
\end{equation}
where the constant $A$ is $A=(\frac{(E+p_x)}{(E-p_x)})$. This
allows us to deal with the $E$ (or $p_x$) equation separately and
for $E$ we get,
\begin{equation}
\frac{dE}{du}=-p_x+\frac{Ep_{x}}{\kappa}=(\frac{1-Ae^{-2u}}{1+Ae^{-2u}})(E-\frac{E^2}{\kappa}).
\label{l5}
\end{equation}
The integration is trivial and we recover the finite Lorentz
transformation for $E$ in $\kappa$-Minkowski spacetime:
\begin{equation}
E'=\frac{\gamma (E-vp_x)}{\alpha}~;~~\alpha
=1+\frac{1}{\kappa}\{((\gamma -1)E-v\gamma p_x\}). \label{l6}
\end{equation}
In the above we define,
$$cosh(u)=\frac{1}{\sqrt{1-v^2}}\equiv \gamma~;~~sinh(u)=v\gamma~;~~e^u=\frac{1+v}{\sqrt{1-v^2}}=\gamma (1+v),$$ where $v$ is the velocity of the primed frame with respect to the unprimed one.  The relation (\ref{l4}) yields the $p_x$ transformation law,
\begin{equation}
p'_x=\frac{\gamma (p_x-vE)}{\alpha}. \label{l7}
\end{equation}
For the transformation law for the transverse momentum $p_y$ we
start from
\begin{equation}
\frac{dp'_y}{du}=\frac{p'_yp'_x}{\kappa}=-\frac{1}{\alpha}\frac{d\alpha
}{du}p'_y, \label{l8}
\end{equation}
where we have exploited the useful identity $$p'_x=-
\frac{\kappa}{\alpha }\frac{d\alpha }{du}.$$ This is integrated to
yield,
\begin{equation}
p'_y=\frac{p_y}{\alpha}. \label{l9}
\end{equation}
We collecting the momentum transformation laws together:
\begin{equation}
E'=\frac{\gamma (E-vp_x)}{\alpha};~p'_x=\frac{\gamma
(p_x-vE)}{\alpha};~p'_y=\frac{p_y}{\alpha};~p'_z=\frac{p_z}{\alpha}.
\label{l10}
\end{equation}
These transformations have been derived before in \cite{kim} from
different considerations and we have based our analysis completely
on the NC $\kappa$-Minkowski phase space. Our method was used in
similar context by \cite{bru} but with a different representation
of $\kappa$-Minkowski phase space algebra.

Before proceeding to derive the $\kappa$-LT for the coordinates
$x_\mu $, let us first find out the new dispersion law that is
$\kappa$-LT invariant. Scanning the following infinitesimal
transformation rules,
\begin{equation}
\{\frac{1}{2}J_{\mu\nu},p^2\}=\frac{p^2}{\kappa}(\eta _\mu p_\nu
-\eta _\nu p_\mu );~~\{\frac{1}{2}J_{\mu\nu},(\eta
p)\}=-(1-\frac{(\eta p)}{\kappa})(\eta _\mu p_\nu -\eta _\nu p_\mu
);~~ \label{l11}
\end{equation}
we find the following combination to be invariant:
\begin{equation}
\{\frac{1}{2}J_{\mu\nu},\frac{p^2}{(1-\frac{(\eta
p)}{\kappa})^2}\}=0. \label{l12}
\end{equation}
The finite $\kappa$-LTs also yields
\begin{equation}
(p^2-m^2(1-\frac{(\eta p)}{\kappa})^2)'=\frac{1}{\alpha
^2}(p^2-m^2(1-\frac{(\eta p)}{\kappa})^2), \label{l13}
\end{equation}
confirming that the new $\kappa$-LT invariant dispersion law is
\begin{equation}
p^2=m^2(1-\frac{(\eta p)}{\kappa})^2. \label{l14}
\end{equation}
This is the MS dispersion law \cite{mag} mentioned at the
beginning (\ref{m}). We recover all the conventional relations in
the $\kappa \rightarrow \infty $ limit.

Now we discuss the generalized spacetime transformation relations.
We follow the same procedure as before and starting from
(\ref{52}) obtain the differential equations for $t$ and $x$,
\begin{equation}
\frac{dt'}{dv}=-x'-\frac{t'p'_x}{\kappa}=-x'+\frac{t'}{\alpha
}\frac{d\alpha
}{dv};~~\frac{dx'}{dv}=-t'-\frac{x'p'_x}{\kappa}=-t'+\frac{x'}{\alpha
}\frac{d\alpha }{dv}, \label{l15}
\end{equation}
which are reexpressed as
\begin{equation}
\frac{d}{dv}(\frac{t'}{\alpha })=-\frac{x'}{\alpha
};~~\frac{d}{dv}(\frac{x'}{\alpha })=-\frac{t'}{\alpha }.
\label{l16}
\end{equation}
A further differentiation separates the variables and we get,
\begin{equation}
\frac{d^2}{dv^2}(\frac{x'}{\alpha })=\frac{x'}{\alpha
};~~\frac{d^2}{dv^2}(\frac{t'}{\alpha })=\frac{t'}{\alpha },
\label{l17}
\end{equation}
giving rise to the solutions,
\begin{equation}
x'=\alpha \gamma (x-vt),~~t'=\alpha \gamma (t-vx), \label{l18}
\end{equation}
where the initial and final conditions  respectively are $x,t$ for
$v=0$ and $x',t'$.  For the transverse coordinate $y$, the
differential equation,
\begin{equation}
\frac{dy'}{dv}=-\frac{y'p'_x}{\kappa}=\frac{y'}{\alpha
}\frac{d\alpha }{dv}, \label{l19}
\end{equation}
induces the transformation,
\begin{equation}
y'=\alpha y. \label{l20}
\end{equation}
Hence  the $\kappa$-LTs for spacetime coordinates  appear as,
\begin{equation}
t'=\alpha \gamma (t-vx);~~x'=\alpha \gamma (x-vt),~~y'=\alpha
y,~~z'=\alpha z. \label{l21}
\end{equation}
These $\kappa$-LTs are precisely the ones derived in \cite{kim} in
a different way where imposing the invariance of $x^\mu p_\mu $
was the starting point and partially derived in \cite{mig} in a Hamiltonian sysplectic framework. In the present paper this derivation is
more systematic and rests solely on the new form of
$\kappa$-algebra (\ref{03}) introduced here.

As in the dispersion relation, once again we look for an invariant
quantity that will generalize the conventional distance and we
find that under the $\kappa$-LT (\ref{l21}),
\begin{equation}
(x^2(1-\frac{(\eta p)}{\kappa})^2)'=x^2(1-\frac{(\eta
p)}{\kappa})^2. \label{l22}
\end{equation}
Hence the invariant length $\mid S \mid$ is generalized to
\begin{equation}
s^2=x^2(1-\frac{(\eta p)}{\kappa})^2. \label{k1}
\end{equation}
This is one of the important results of the present paper. Its $1+1$-dimensional analogue was suggested in \cite{mig}. \vskip
.4cm
{\bf{III: Canonical Variables}}\\
\vskip .2cm In this section we will introduce a new set of phase
space variables which obey canonical Poisson brackets are
transform in the conventional way under SR Lorentz transformation.
Somewhat similar considerations in parts have appeared before in
\cite{mig,der,hos} but exhaustive study of the {\it{full}} canonical
phase space as presented here is new. Indeed,  these variables are
composites of phase space coordinates will have to suitable
ordered upon quantization. But, in the classical framework they
will prove to be very convenient and they drastically simplify the
computations while analyzing phenomenological consequences of the
modified Lorentz transformations. We will return to the quantum
case at the end.

The two invariant quantities that we derived in (\ref{m},\ref{k1})
suggest the forms of these canonical avatars:
\begin{equation}
X_\mu \equiv x_\mu (1-\frac{(\eta p)}{\kappa})=x_\mu
(1-\frac{E}{\kappa});~~P_\mu \equiv \frac{p_\mu}{(1-\frac{(\eta
p)}{\kappa})}=\frac{p_\mu}{(1-\frac{E}{\kappa})}. \label{c1}
\end{equation}
We remind  that the variables on the right hand side obey
$\kappa$-LT laws. Using the NC algebra (\ref{03}) it is easy to
check the following:
\begin{equation}
\{X_\mu ,P_\nu \}=-g_{\mu\nu};~~\{X_\mu ,X_\nu \}=\{P_\mu ,P_\nu
\}=0. \label{c2}
\end{equation}
Hence the $X,P$ phase space is canonical. The above relations in
(\ref{c1}) are invertible,
\begin{equation}
x_\mu = X_\mu (1+\frac{(\eta P)}{\kappa})=X_\mu
(1+\frac{P_0}{\kappa});~~p_\mu = \frac{P_\mu}{(1+\frac{(\eta
P)}{\kappa})}=\frac{P_\mu}{(1+\frac{P_0}{\kappa})}. \label{c3}
\end{equation}
Next we consider Lorentz transformations of the canonical
variables and find, for example,
$$T'=t'(1-\frac{E}{\kappa})'=\gamma \bar \alpha (t-vx)[1-\frac{\gamma}{\kappa \bar \alpha}(E-vp_x)]$$
\begin{equation}
=\gamma [t(1-\frac{E}{\kappa})-vx(1-\frac{E}{\kappa})]=\gamma
(T-vX), \label{c4}
\end{equation}
where $\bar \alpha =\alpha (-v)$ and we have used the identity
$(\bar \alpha )^{-1}=\alpha '$. In an identical fashion we can show,
\begin{equation}
X'=x'(1-\frac{E}{\kappa})'=\gamma
(X-vT);~~Y'=y'(1-\frac{E}{\kappa})'=Y;~~Z'=Z. \label{c5}
\end{equation}
Exploiting similar considerations we also ascertain that,
$$
P'_0=\frac{E'}{(1-\frac{E}{\kappa})'}=\gamma
(P_0-vP_x);~~P'_x=\frac{p'_x}{(1-\frac{E}{\kappa})'}=\gamma
(P_x-vP_0);$$
\begin{equation}
P'_y=\frac{p'_y}{(1-\frac{E}{\kappa})'}=P_y,~~P'_z=P_z. \label{c6}
\end{equation}
To further convince ourselves about the validity of the canonical
variable approach let us study the group property of these new
Lorentz transformations. Consider, for example, two successive
$\kappa$-LT' on $x$,
$$x=\alpha '_1(C_1x'+S_1t');~~t=\alpha '_1(C_1t'+S_1x'),$$
\begin{equation}
x'=\alpha ''_2(C_2x''+S_2t'');~~t'=\alpha ''_2(C_2t''+S_2x''),
\label{c7}
\end{equation}
where $$C_1=cosh(u_1)=\frac{1}{\sqrt{1-v^{2}_{1}}}=\gamma
_1;~S_1=sinh(u_1)=v_1\gamma
_1;~\alpha'_1=1-\frac{1}{\kappa}\{(1-C_1)E'-S_1p'_x\}$$ and so on.
We will also require the energy-momentum $\kappa$-LTs,
\begin{equation}
E'=\frac{1}{\alpha ''_2}(C_2E''+S_2p''_x);~~p'_x=\frac{1}{\alpha
''_2}(C_2p''_x+S_2E''). \label{c8}
\end{equation}
Concentrating on the $x$-transformation, the above will lead to,
\begin{equation}
x=\alpha '_1[C_1\{\alpha ''_2(C_2x''+S_2t'')\}+S_1\{\alpha
''_2(C_2t''+S_2x'')\}]=\alpha
''_{(1+2)}[C_{(1+2)}x''+S_{(1+2)}t''], \label{c9}
\end{equation}
where $(1+2)$ stands for $(v_1+v_2)$. However, in terms of the
canonical variables we should simply obtain,
\begin{equation}
X=C_{(1+2)}X''+S_{(1+2)}T'', \label{c10}
\end{equation}
which, when expressed in terms of NC physical variables yields,
$$
(1-\frac{E}{\kappa})x=C_{(1+2)}(1-\frac{E''}{\kappa})x''+S_{(1+2)}(1-\frac{E''}{\kappa})t''$$
\begin{equation}
\Rightarrow
x=\frac{(1-\frac{E''}{\kappa})}{(1-\frac{E}{\kappa})}C_{(1+2)}x''+S_{(1+2)}t''=\alpha''_{(1+2)}(C_{(1+2)}x''+S_{(1+2)}t'').
\label{c11}
\end{equation}
This is same as (\ref{c9}). Hence the group property is
established for $\kappa$-LTs and also the fact that the canonical
$X_\mu ,P_\mu $ variables obey SR LTs.

The important lesson that we learn from the above discussion is
that the $(X_\mu ,P_\mu )$ phase space is truly canonical in the
sense that it satisfies canonical Poison brackets and moreover
obeys normal $(\kappa \rightarrow \infty )$ SR LTs. Hence, {\it{as
far as classical physics is concerned}}, we can directly borrow
the  SR kinematical laws (by writing them in terms of $(X_\mu
,P_\mu )$, and subsequently deduce the laws in $\kappa$-NC
spacetime by rewriting them in terms of $(x_\mu ,p_\mu )$, using
the mapping (\ref{c1}). We will put this idea in immediate use in the next section where we obtain the deformed symmetry generators. However, as we will discuss in later
sections, one has to be more careful while establishing dynamical
laws, even though the basic approach remains the same. In the last
section, we will briefly comment on the quantization problem,
which is very interesting. 
\vskip .4cm
{\bf{IV: Deformed Symmetry Generators}}\\
\vskip .2cm 

Let us understand  the fate of the conventional symmetry
principles in the $\kappa$-extended particle model. We will see
that we need deformed  symmetry generators.

In the conventional case, the phase space algebraic structure of
the point particle is invariant under the following symmetry
transformations: translation, Lorentz rotation, dilation and
special conformal transformation. On the other hand, the particle
dispersion relation $P^2-m^2=0$ enjoys invariance under
translation and Lorentz rotation, and the mass term $m$ breaks the
symmetry under dilation and special conformal transformation.
Finally, the symmetry generators satisfy a closed algebra among
themselves.

In the $\kappa$-particle model our aim is to construct the
generators in the $\kappa$-NC space that preserve invariances of
both the $\kappa$-NC phase space algebra (\ref{03})  and the
structure of the algebra among generators (see below in (\ref{0s})). Then we will
check how the $\kappa$-modified dispersion relation (MS relation
(\ref{m}) in the present case) is affected. Once again the
canonical $(X_\mu ,P_\mu )$ variables will do the trick. The idea
is to first write down the generators in terms of $(X_\mu ,P_\mu
)$ degrees of freedom using the conventional form of the
generators, ({\it{i.e.}} that of normal particle in normal phase
space). They will obviously satisfy the standard closed algebra among
generators:
$$
\{J^{\mu\nu },J^{\alpha\beta }\}=g^{\mu\beta }J^{\nu\alpha
}+g^{\mu\alpha }J^{\beta \nu}+g^{\nu\beta }J^{\alpha\mu
}+g^{\nu\alpha }J^{\mu\beta }~;~~\{J^{\mu\nu },T^\sigma \}=g^{\nu\sigma}T^{\mu}-g^{\mu\sigma}T^{\nu}~; $$
$$ \{J^{\mu\nu },D \}=0~;~~ \{J^{\mu\nu },K^\sigma \}=2D(g^{\nu\sigma}X^{\mu}-g^{\mu\sigma}X^{\nu})-X^2(g^{\nu\sigma}T^{\mu}-g^{\mu\sigma}T^{\nu})~;$$
$$\{T^\mu ,T^\nu \}=0~;~~\{T^\mu ,D\}=T^\mu ~;~~\{T^\mu ,K^\nu \}=2Dg^{\mu\nu}-2J^{\mu\nu}~;$$
\begin{equation}
\{D ,D \}=0~;~~\{D,K^\mu \}=K^\mu ~;~~\{K^\mu ,K^\nu \}=0~,
 \label{0s}
\end{equation}
where $J_{\mu\nu}~,T_\mu~,~D$ and $K_\mu $ stand for generators of
Lorentz rotation, translation, dilation and special conformal
transformation respectively. Their structures are given by,
$$
J_{\mu\nu}=X_\mu P_\nu -X_\nu P_\mu~;~~T_\mu = P_\mu ~;~~D=(XP)~;
$$
\begin{equation}
K_\mu =2(XP)X_\mu -X^2P_\mu~.
 \label{1s}
\end{equation}
Next we exploit the map $(X_\mu ,P_\mu )\rightarrow (x_\mu ,p_\mu
)$ given in (\ref{c1}) to rewrite the generators in the $\kappa$-NC
spacetime:
$$
j_{\mu\nu}=x_\mu p_\nu-x_\nu p_\mu ~;~~ t_\mu
=\frac{p_\mu}{1-(\eta p)/\kappa }~;~~~;~~d=(xp);$$
\begin{equation}
k_\mu =(1-(\eta p)/\kappa)[2(xp)x_\mu -x^2p_\mu ]~. \label{s2}
\end{equation}
{\it{By construction, the generators in (\ref{s2}) will satisfy
the same algebra (\ref{0s}) provided one uses the $\kappa$-NC
algebra (\ref{03})}}. These are the deformed 
generators. The infinitesimal transformation operators are,
$$
j=\frac{1}{2}a^{\mu\nu}j_{\mu\nu}=\frac{1}{2}a^{\mu\nu}(x_\mu
p_\nu-x_\nu p_\mu )~;~~ t=a^\mu t_\mu =\frac{(ap)}{1-(\eta
p)/\kappa }~;~~d=a(xp)~;$$
\begin{equation}
k=a^\mu k_\mu =(1-(\eta p)/\kappa)[2(xp)(ax) -x^2(ap) ],
\label{s3}
\end{equation}
where generically $a$ denotes the infinitesimal parameter. Using
the definition of small change in $A$ due to transformation
$\delta _b$ as,
\begin{equation}
\delta _b A=\{\delta _b,A\},
 \label{12s}
\end{equation}
we compute the explicit forms of transformations:
\begin{equation}
\delta _j x_\mu =a^{\alpha\beta}(g_{\beta\mu}x_\alpha
+\frac{1}{\kappa}p_\alpha\eta _\beta x_\mu )~;~~ \delta _j p_\mu
=a^{\alpha\beta}(g_{\beta\mu}p_\alpha
-\frac{1}{\kappa}p_\alpha\eta _\beta p_\mu ),\label{s4}
\end{equation}
\begin{equation}
\delta _t x_\mu =\frac{a_\mu}{1-(\eta p)/\kappa }~;~~\delta _t p_\mu =0,
\label{s5}
\end{equation}
\begin{equation}
\delta_dx_\mu = a(1-(\eta p)/\kappa)x_\mu ~;~~\delta_dp_\mu =-
a(1-(\eta p)/\kappa)p_\mu , \label{s6}
\end{equation}
$$
\delta _kx_\mu =(1-(\eta p)/\kappa)[2(ax)x_\mu -x^2a_\mu
+\frac{2}{\kappa}x_\mu (-(xp)(a\eta )-(\eta p)(ax)+(\eta
x)(ap))]~;$$
\begin{equation}
\delta _kp_\mu =2(1-(\eta p)/\kappa)[((ap)x_\mu -(ax)p_\mu
-(xp)a_\mu ) +\frac{1}{\kappa}p_\mu ((xp)(a\eta )+(\eta
p)(ax)-(\eta x)(ap))].
 \label{s7}
\end{equation}
Clearly the variations differ from their commutative spacetime
counterpart. Next we want to ascertain that the $\kappa$-NC
algebra (\ref{03}) is stable under the above symmetry operations.
This is done by checking the validity of the identity,
\begin{equation}
\{A,B\}=C ~\Rightarrow \delta_b \{A,B\}=\delta _bC ,\label{s8}
\end{equation}
or more explicitly,
\begin{equation}
\{\delta_bA,B\}+\{A,\delta_bB\}=\delta _bC .\label{s9}
\end{equation}
In the above we refer to (\ref{03}) for $\{A,B\}=C$ and (\ref{s4}-\ref{s7}) for $\delta_{b}$.
A straightforward but tedious calculation shows that the above
identity is, indeed, valid. This assures us about the consistency
of the whole procedure.

Regarding the behavior of the MS dispersion law (\ref{m}) we have
already checked that it is $\kappa$-Lorentz invariant. It is also
trivially translation invariant. The variation under dilatation is
given by,
\begin{equation}
\delta_d(p^2-m^2(1-(\eta p)/\kappa)^2)=-2(1-(\eta
p)/\kappa)[p^2+m^2(1-(\eta p)/\kappa)\frac{(\eta
p)}{\kappa}]=-2p^2, \label{s10}
\end{equation}
where the MS relation is imposed on the right hand side.
In fact this variation is structurally identical to that of the
normal particle showing that in $\kappa$-NC spacetime also the
mass terms breaks dilation invariance in the same way.

For  special conformal transformation, the variation is given by,
$$
\delta_k (p^2-m^2(1-(\eta p)/\kappa)^2)=2(1-(\eta
p)/\kappa)[-(ax)p^2 +\frac{1}{\kappa}(p^2-m^2(1-(\eta
p)/\kappa)^2) ((xp)(a\eta )$$
\begin{equation}
+(\eta p)(ax)-(\eta x)(ap))].
\label{s11}
\end{equation}
If we impose the MS law on the right hand side of (\ref{s11}) we
find the variation to be,
\begin{equation}
\delta_k (p^2-m^2(1-(\eta p)/\kappa)^2)=-2(1-(\eta
p)/\kappa)(ax)p^2 ,\label{s12}
\end{equation}
which very clearly mimics the normal particle characteristics.

The above results indicate how the $\kappa$-extension modifies the
mathematical structure of the normal (Special Theory) relativistic
particle. At the same time we have explicitly constructed the full set of deformed  symmetry generators in $\kappa$-spacetime that have all the attributes of these generators in normal comutative spacetime.

\vskip .4cm
{\bf{V: $\kappa$-Lorentz Transformation Phenomenology}}\\
\vskip .2cm Let us start by discussing an important issue: the
$\kappa $-LT of velocity, that will lead to the velocity  addition
theorem. In our canonical setup, the three velocity (in the primed
frame) can be defined in the conventional way,
\begin{equation}
\vec U'=U'_X\hat i+U'_Y\hat j+U'_Z\hat
k;~~U'_X=\frac{dX'}{dT'},...  . \label{p1}
\end{equation}
Hence, in terms of these variables, the velocity  (in an unprimed
frame, moving with velocity $v$ in $X$-direction) will obey
conventional Special Theory relations,
$$
\vec W=W_X\hat i+W_Y\hat j+W_Z\hat k;~~W_X=\frac{dX}{dT},..$$
\begin{equation}
W_X=\frac{U'_X+v}{1+U'_Xv};~~W_X=\frac{U'_X+v}{1+U'_Xv};~~W_Y=\frac{U'_Y}{1+U'_Xv};~~W_Z=\frac{U'_Z}{1+U'_Xv}.
\label{p2}
\end{equation}
Let us now map this relation to NC $(x_\mu ,p_\mu )$ space. From
the mapping (\ref{c1}) it is clear that the velocity components
are unchanged,
\begin{equation}
\frac{dX}{dT}=\frac{dx}{dt}, ...\frac{dX'}{dT'}=\frac{dx'}{dt'},
.... \label{p3}
\end{equation}
which means that the velocity transformation laws will not change
in the $\kappa$-spacetime and in particular the velocity addition
law will remain unchanged.

However, in reality the analysis is more tricky and so far as we
have discussed, it validity is restricted only to {\it{free}}
$\kappa $-particle. In fact this statement also needs to be proved
which we do in Section IV by considering a specific model for a
$\kappa$-particle. To understand the problem let us follow the
derivation of the velocity addition rule more closely. From
$\kappa$-LT the velocity components are related by,
$$
x=\gamma \alpha '(x'+vt')$$
\begin{equation}
\Rightarrow \frac{dx}{dt}=\gamma \alpha
'\frac{dt'}{dt}(\frac{dx'}{dt'}+v)+\psi \Rightarrow w_x=\gamma
\alpha '\frac{dt'}{dt}(u'_x+v)+\psi . \label{p4}
\end{equation}
The piece $\psi $ can come from $(dE)/(dt)$  if the particle is
interacting. Note that the $(dE)/(dt)$ term will contribute to
$O(1/\kappa )$. For free particles we will show in Section V that
$(dE)/(dt)=0$ and so it will not affect  (\ref{p4}). Hence,
for free particle we have,
\begin{equation}
w_x=\gamma \alpha '\frac{dt'}{dt}(u'_x+v).
\label{p5}
\end{equation}
In a similar way, for the free particle we will have,
\begin{equation}
\frac{dt'}{dt}=\frac{\gamma}{\alpha '}(1-vw_x), \label{p6}
\end{equation}
where we have  taken $p_x$ to be time independent, which we prove
later. Substituting (\ref{p6}) in (\ref{p5}) we get
\begin{equation}
w_x=\frac{u'_x+v}{1+u'_xv}. \label{p7}
\end{equation}
Similar results will follow for other velocity components.

In case of $\kappa$-LT for acceleration components $(a_x)$ for the
free particle we find
\begin{equation}
a_x=\frac{1}{\alpha '}\frac{a'_x}{(\gamma
^2(1+u'_xv))^3}. \label{p8}
\end{equation}
Hence the $\kappa$-LT law for acceleration  is changed even for
the free particle, in contrast to the velocity relation, which
remains unaltered. The $\kappa $-modification for the other
components of accleration are same in nature.

From the structure of $\kappa$-LT invariants (\ref{l14},\ref{l22})
and the mappings (\ref{c1}), it is clear that the phase space
variables $x_\mu $ and $p_\mu $ get mixed up in an inseparable
way. This is  the reason why kinematic laws are not isolated
from the dynamical behavior of the particle. In case the particle
is not free, the simplest possibility is that it is under the
influence of an external force $\vec F$. In that case we should
use
\begin{equation}
\frac{dE}{dt}=\vec F.\vec w , \label{p9}
\end{equation}
and so there will $O(\frac{1}{\kappa })$ corrections, depending on
the force, in all the relations.

We can also study Lorentz contraction. We find that
\begin{equation}
X_2-X_1=\gamma (X'_2-X'_1)\Rightarrow x_2-x_1=\gamma  (\alpha
'_2x'_2-\alpha '_1x'_1), \label{p10}
\end{equation}
where $2$ and $1$ refer to two spacetime positions and $\alpha _2$
and $\alpha _1$ are the values of $\alpha$ at positions $2$ and
$1$ respectively. Once again for a free particle we use,
\begin{equation}
E_2=E_1~;~~(p_x)_2=(p_x)_1~, \label{p11}
\end{equation}
and  find the modified Lorentz contraction law for length $l$ as,
\begin{equation}
l'=\frac{l}{\gamma}[1-\frac{1}{\kappa}\{(1-\gamma )E+v\gamma p_x
\}] ~, \label{p12}
\end{equation}
applicable to free particle.

It is curious to see that in the normal $(\kappa \rightarrow
\infty )$ case,
$$l'=\frac{l}{\gamma}\approx l(1-\frac{v^2}{2})+O(v^3),$$
and in the $\kappa$-spacetime, keeping terms up to $O(v)$ in the
non-relativistic limit, we have,
$$l'\approx l(1-\frac{vp_x}{\kappa})\sim l(1-\frac{mv^2}{\kappa})$$ for $p_x\sim mv$.

In the next section we will formulate an explicit model of a point
particle that has a $\kappa$-Minkowski NC phase space (\ref{03})
and Magueijo-Smolin dispersion law (\ref{l14}). This will help us
to understand how,  some of the assumptions that we have made in
this section, regarding the particles energy and momentum, are
concretely realized. \vskip .4cm
{\bf{VI: Lagrangian for $\kappa$-Particle }}\\
\vskip .2cm In this section we construct a Lagrangian for the
$\kappa$-particle. This has been a topic of recent interest and
several authors \cite{others,sg1}  have proposed models for
particles with NC phase space of different structures. However,
the model we propose here for $\kappa$-NC phase space is quite
elegant and can be expressed in a closed form.

 Again the canonical variable approach becomes
convenient since we are sure that the relativistic free particle
action in terms of canonical $(X_\mu ,P_\mu )$ degrees of freedom
will be,
\begin{equation}
L=(P^\mu\dot X_\mu)-\lambda (P^2-m^2). \label{a1}
\end{equation}
We now convert this $L$ to a function depending on physical
$\kappa $-NC phase space coordinates:
$$
L=(\frac{p^\mu}{1-\frac{(\eta p)}{\kappa }})(x_\mu (1-\frac{\eta
p)}{\kappa}))^.-\frac{\lambda}{2}(\frac{p^2}{(1-\frac{(\eta
p)}{\kappa})^2}-m^2)$$
\begin{equation}
=(p\dot x)-\frac{(px)(\eta \dot p)}{\kappa (1-\frac{(\eta
p)}{\kappa})}-\frac{\lambda}{2}(p^2-m^2(1-\frac{(\eta
p)}{\kappa})^2), \label{a2}
\end{equation}
where we have redefined the arbitrary multiplier $\lambda $. Our
claim is that the symplectic structure in (\ref{a2}) will induce
the $\kappa$-NC phase space algebra and the $\lambda $-term will
obviously impose the MS mass shell condition. A slightly differenct form of symplectic structure in $1+1$-dimensions is given in \cite{mig}. We now proceed to
demonstrate the former in Dirac's Hamiltonian constraint analysis
scheme \cite{dirac}.

In the above first order Lagrangian (\ref{a2}) $x_\mu $ and $p_\mu
$ are treated as independent variables. The conjugate momenta are,
\begin{equation}
\pi^{x}_{\mu}=\frac{\partial L}{\partial \dot x^\mu}=p_\mu
~;~~\pi^{p}_{\mu}=\frac{\partial L}{\partial \dot
p^\mu}=-\frac{(px)}{\kappa (1-\frac{(\eta p)}{\kappa})}\eta _\mu .
\label{a3}
\end{equation}
The only nontrivial Poisson brackets are,
\begin{equation}
\{x_\mu ,\pi^{x}_{\nu}\}=\{p_\mu ,\pi^{p}_{\nu}\}=-g_{\mu\nu}.
\label{a4}
\end{equation}
The momenta equation (\ref{a3}) shows that there are two sets of
constraints:
\begin{equation}
\psi^{1}_{\mu}\equiv \pi^{x}_{\mu}-p_\mu \approx
0~;~~\psi^{2}_{\mu}\equiv \pi^{p}_{\mu}+\frac{(px)}{\kappa
(1-\frac{(\eta p)}{\kappa})}\eta _\mu \approx 0. \label{a5}
\end{equation}
In the terminology of Dirac constraint analysis \cite{dirac}, the
non-commutating constraints are termed as Second Class Constraints
(SCC) and the commutating constraints, that induce local gauge
invariance, are First Class Constraints (FCC). In the presence of
SCCs $(\psi^{1},\psi^{2})$ that do not commute (as Poisson
Brackets), $\{\psi^1 ,\psi ^2 \}\ne 0$, the modified symplectic
structure (or Dirac Brackets) are defined in the following way,
\begin{equation}
\{A,B\}^*=\{A,B\}-\{A,\psi ^i\}\{\psi ^i,\psi ^j\}^{-1}\{\psi
^j,B\}, \label{a6}
\end{equation}
where $\{\psi ^i,\psi ^j\}$ refers to the constraint matrix.

For the set of constraints (\ref{a5}) the constraint matrix is
\begin{equation}
\{\psi ^i_\mu,\psi ^j_\nu\}=
 \left (
\begin{array}{cc}
0  & g_{\mu\nu}+\frac{p_\mu \eta_\nu}{\kappa (1-\frac{(\eta p)}{\kappa})} \\
-g_{\mu\nu}-\frac{p_\nu \eta_\mu}{\kappa (1-\frac{(\eta
p)}{\kappa})} &  \frac{x_\mu \eta_{\nu}-x_\nu \eta_{\mu}}{\kappa
(1-\frac{(\eta p)}{\kappa})}
\end{array}
\right ) . \label{a7}
\end{equation}
The inverse matrix is computed to be,
\begin{equation}
\{\psi ^i_\nu,\psi ^j_\lambda\}^{-1}=
 \left (
\begin{array}{cc}
\frac{1}{\kappa}(x_\nu \eta_{\lambda}-x_\lambda \eta_\nu) & -g_{\nu\lambda}+\frac{1}{\kappa}\eta_\nu p_\lambda  \\
g_{\nu\lambda}-\frac{1}{\kappa}\eta_\lambda p_\nu  &  0
\end{array}
\right ) . \label{aa7}
\end{equation}
Exploiting the definition (\ref{a6}) we obtain the Dirac Brackets,
\begin{equation}
\{x_\mu ,x_\nu \}=\frac{1}{\kappa}(x_\mu \eta_{\nu}-x_\nu
\eta_{\mu })~;~~
\{x_{\mu},p_{\nu}\}=-g_{\mu\nu}+\frac{1}{\kappa}\eta_{\mu}p_{\nu}~;~~\{p_{\mu},p_{\nu}\}=0,
 \label{a8}
\end{equation}
which is nothing but the $\kappa $-NC phase space introduced at
the beginning (\ref{03}). Hence $L$ in (\ref{a2}) correctly
reproduces the $\kappa$-NC symplectic structure.

It is also very useful to have the Nambu-Goto form of action
comprising only of coordinate space variables for the $\kappa
$-particle. This can be recovered  by eliminating $p_\mu $ and
$\lambda $ from  $L$ given in  (\ref{a2}), which we now proceed to
do.

The variational equations of motion from (\ref{a2}) are,
\begin{equation}
\dot x_\mu -\frac{(\eta \dot p)}{\kappa (1-\frac{(\eta
p)}{\kappa})}x_\mu +\frac{(px)^.}{\kappa (1-\frac{(\eta
p)}{\kappa})}\eta _\mu -\lambda (p_\mu
+\frac{m^2}{\kappa}(1-\frac{(\eta p)}{\kappa})\eta_{\mu})=0,
\label{a9}
\end{equation}
\begin{equation}
\dot p_\mu +\frac{(\eta \dot p)}{\kappa (1-\frac{(\eta
p)}{\kappa})}p_\mu =0. \label{a10}
\end{equation}
Rewriting (\ref{a10}) in the form,
$$\dot p_\sigma (g^{\sigma\mu}+\frac{\eta^{\sigma p^{\mu}}}{\kappa (1-\frac{(\eta p)}{\kappa})})\equiv \dot p_\sigma G^{\sigma\mu}=0,$$
we observe that $G^{-1}$ exists and hence (\ref{a10}) yields,
\begin{equation}
\dot p_\mu =0. \label{a11}
\end{equation}
This condition considerably simplifies (\ref{a9}) to,
\begin{equation}
\dot x_\mu  +\frac{(p \dot x)}{\kappa (1-\frac{(\eta
p)}{\kappa})}\eta _\mu =\lambda (p_\mu
+\frac{m^2}{\kappa}(1-\frac{(\eta
p)}{\kappa})\eta_{\mu})\Rightarrow (1-\frac{(\eta
p)}{\kappa})=\frac{1}{\kappa}\sqrt{\frac{(p\dot x)}{\lambda}}.
\label{a12}
\end{equation}
We find the following solutions,
\begin{equation}
p_\mu =\frac{\dot x_\mu}{\lambda} ~;~~\lambda =\frac{\sqrt{\dot
x^2}}{m}+\frac{1}{\kappa}(\eta \dot x). \label{a13}
\end{equation}
Consistency of the computation is checked by noting that the
solutions (\ref{a13}) satisfies the correct dispersion law we have
imposed,
\begin{equation}
p^2=\frac{(p\dot x)}{\lambda}=m^2(1-\frac{(\eta p)}{\kappa})^2.
\label{a14}
\end{equation}
Finally we obtain the cherished Nambu-Goto Lagrangian for the
$\kappa $-particle:
\begin{equation}
L=\frac{m\sqrt{\dot x^2}}{(1+\frac{m(\eta \dot x)}{\kappa
\sqrt{\dot x^2}})}(1+\frac{m}{\kappa}(\eta \dot x)(\frac{(x\dot
x)}{\sqrt{\dot x^2}})^.). \label{a15}
\end{equation}
Notice that (\ref{a15}) is a higher derivative Lagrangian.
Hamiltonian analysis of it will yield the $\kappa $-NC phase space
algebra.

The equation of motion $\dot p_\mu =0$ in (\ref{a11}) is very
important. It is the analogue of the free particle equation of
motion in normal spacetime (Newton's law) showing that momentum is conserved for
the free $\kappa $-particle as well. We have already exploited
this condition in Section III. \vskip .4cm
{\bf{VII: Quantization}}\\
\vskip .2cm In this brief section we will try to argue through a
few applications that the canonical coordinate approach is very
convenient for quantization of the $\kappa $-particle. We recall
from (\ref{c3})  the mapping between NC $\kappa $-coordinates
$(x_\mu, p_\mu )$ and canonical coordinates $(X_\mu, P_\mu )$:
$$x_\mu = X_\mu (1+\frac{(\eta P)}{\kappa});~~p_\mu = \frac{P_\mu}{(1+\frac{(\eta P)}{\kappa})}.$$
It is straightforward to elevate the classical Poisson brackets in
$(X_\mu, P_\mu )$ to quantum commutation relations,
\begin{equation}
[P_\mu ,X_\nu]=ig_{\mu\nu}~;~~[X_\mu ,X_\nu]=[P_\mu ,P_\nu]=0.
\label{q}
\end{equation}
In quantum theory the mapping has to be between operators,
\begin{equation}
\hat x_\mu = \hat X_\mu (1+\frac{\hat P_0}{\kappa});~~\hat p_\mu =
\frac{\hat P_\mu}{(1+\frac{\hat P_0}{\kappa})}. \label{q1}
\end{equation}
However, notice that only the mapping between $x_0\equiv t$ and
$X_0\equiv T$ needs operator ordering. Clearly there is no
operator ordering ambiguity in non-relativistic quantum mechanical
problems and since $P_0$ commutes with all the relevant variables
$\vec P$ and $\vec X $ and one can replace $P_0$ by a number $m$,
the particle mass in the non-relativistic limit. Hence the mapping
becomes,
\begin{equation}
\hat x_i = \hat X_i (1+\frac{(m)}{\kappa});~~\hat p_\mu =
\frac{\hat P_\mu}{(1+\frac{m}{\kappa})}. \label{q2}
\end{equation}
The advantage is that the operators $\hat P_i$ and $X_i$ can be
treated in the conventional way.

Let us use the above ideas in the simplest problem, {\it{i.e.}}
(non-relativistic) harmonic oscillator in $\kappa $-spacetime. The
Hamiltonian is,
$$
H=\frac{\vec p^2}{2m}+\frac{K}{2}x^2 $$
\begin{equation}
 =\frac{\vec P^2}{2m(1+\frac{(m)}{\kappa})^2}+\frac{K}{2}(1+\frac{(m)}{\kappa})^2X^2 \equiv \frac{\vec P^2}{2\tilde m}+\frac{\tilde K}{2}x^2.
\label{q3}
\end{equation}
This shows that the frequency,
\begin{equation}
\omega =\sqrt{\frac{\tilde K}{\tilde m}}=\sqrt{\frac{ K}{ m}},
\label{q4}
\end{equation}
and hence energy levels will remain unchanged.

We can consider the Hydrogen atom problem where the Hamiltonian
is,
$$H=\frac{\vec p^2}{2m}-\frac{Ze^2}{r}$$
\begin{equation}
=\frac{\vec
P^2}{2m(1+\frac{(m)}{\kappa})^2}-\frac{Ze^2}{R(1+\frac{m}{\kappa})}\equiv
\frac{\vec P^2}{2\tilde m}-\frac{Z\tilde e^2}{R}. \label{q5}
\end{equation}
Once again there is no change in the energy levels. This is
actually obvious since  the only effect of $\kappa $ in
non-relativistic physics turns out to be a numerical scaling of
the operators which is not observable.

\vskip .4cm
{\bf{VIII: Conclusion and Outlook}}\\
\vskip .2cm Let us summarize our work. We have concentrated on a
particular Doubly Special Relativity model that has an underlying
novel form of $\kappa$-Minkowski Noncommutative phase space
structure along with a Magueijo-Smolin form of modified dispersion
law. The novel and attractive feature of our work is the
construction of (invertible) mapping between the Noncommutative
$\kappa$-Minkowski  phase space and a {\it{completely canonical
phase space}} the latter obeying normal Poisson brackets. We
further show the the canonical degrees of freedom transform under
Special Theory Lorentz transformations whereas the original
physical variables transform under $\kappa$-extended Lorentz
transformations. Following this approach we have developed the complete structure of $\kappa$-deformation of the symmetry generators. These modified generators obey an un-deformed algebra, keep the $\kappa$-noncommutative phase space algebra stable under symmetry operations but induce deformed  transformations on the phase space variables. Furthermore, the canonical construction simplifies the study of
particle behavior  in this particular DSR framework tremendously
since the existing results of Special Theory directly applicable
in the artificial canonical space. But from the explicit relation
we have provided, these results can be mapped to the physical
$\kappa$-Minkowski phase space, the latter having a much more
complex Noncommutative Hamiltonian structure and generalized
Lorentz transformations. Using our scheme, the free
$\kappa$-particle action developed here can be extended to include
interactions in a straightforward way. An outcome of our analysis
is that $\kappa$-effects will probably not be visible in
non-relativistic quantum systems.

As the next step in our programme we plan to investigate two problems:\\
(i) Behavior of $\kappa $-particle in the presence of interactions.
We expect kinematical relations will change as we have indicated.
We will introduce interaction term in the free particle Lagrangian
in the canonical space since in canonical coordinates the interaction
terms will have normal relativistic structure. Subsequently they can be
mapped to $\kappa$-phase space coordinates and the dynamics can be studied.\\
(ii) Construction of field theory  in $\kappa$-NC spacetime. This
is a very important issue and requires immediate attention. Our
aim is to exploit the canonical approach once again. We intend to
consider the field map $\varphi (x)\Rightarrow \phi (X)$. Notice
that field theoretic Poisson brackets among $\varphi (x)$ and its
momentum field will be  non-canonical, (induced by $(x,p)-\kappa
$-NC algebra), but  brackets among  $\phi (X)$ and its conjugate
momentum field will be canonical. In this way it will be possible
to map a field theory in $\kappa$-NC spacetime to a field theory
in normal spacetime with normal fields. This construction should be compared with the (to be constructed) conventional $*$-product formulation of NC field theory.
\vskip .4cm
{\it{Acknowledgements}}: We are grateful to Professor S.Mignemi for informing us about \cite{mig}. Also it is a pleasure to thank Professor S.Meljanac for a number of helpful correspondences.

\end{document}